\documentclass[journal=jacsat,manuscript=article]{achemso}

\usepackage{xcolor}
\usepackage[version=3]{mhchem} 

\DeclareUnicodeCharacter{0301}{\'{e}}

\author{Anastasiia Zalogina}
\affiliation{Nonlinear Physics Center, Research School of Physics, Australian National University, Canberra ACT 2601, Australia}
\author{Pavel Tonkaev}
\affiliation{Nonlinear Physics Center, Research School of Physics, Australian National University, Canberra ACT 2601, Australia}
\author{Aditya Tripathi}
\affiliation{Nonlinear Physics Center, Research School of Physics, Australian National University, Canberra ACT 2601, Australia}
\author{Hoo-Cheol Lee}
\affiliation{Department of Physics, Korea University, Seoul 02841, Republic of Korea}
\author{Luca Carletti}
\affiliation{Department of Information Engineering, University of Brescia, Brescia 25123, Italy} 
\author{Hong-Gyu Park}
\affiliation{Department of Physics, Korea University, Seoul 02841, Republic of Korea}
\author{Sergey S. Kruk}
\email{sergey.kruk@outlook.com}
\affiliation{Nonlinear Physics Center, Research School of Physics, Australian National University, Canberra ACT 2601, Australia}
\alsoaffiliation{Department of Physics, Paderborn University, 33098 Paderborn, Germany}
\author{Yuri Kivshar}
\email{yuri.kivshar@anu.edu.au}
\affiliation{Nonlinear Physics Center, Research School of Physics, Australian National University, Canberra ACT 2601, Australia}

\title[An \textsf{achemso} demo]
  {Enhanced five-photon photoluminescence in subwavelength AlGaAs resonators}

\keywords{\textit{Nanoresonators, subwavelength photonics, multiphoton photoluminescence, Mie resonances}}

\begin{document}








\begin{abstract}
Multiphoton processes of absorption photoluminescence have enabled a wide range of applications including three-dimensional microfabrication, data storage, and biological imaging. While the applications of two-photon and three-photon absorption and luminescence have matured considerably, higher-order photoluminescence processes remain more challenging to study due to their lower efficiency, particularly in subwavelength systems. Here we report the observation of {\it five-photon luminescence} from a single subwavelength nanoantenna at room temperature enabled by the Mie resonances. We excite an AlGaAs resonator at around 3.6~$\mu$m and observe photoluminescence at around 740~nm. We show that the interplay of the Mie multipolar modes at the subwavelength scale can enhance the efficiency of the five-photon luminescence by at least four orders of magnitude, being limited only by sensitivity of our detector. Our work paves the way towards applications of higher-order multiphoton processes at the subwavelength scales enabled by the physics of Mie resonances.

\end{abstract}


\newpage

Multiphoton absorption (MPA) and photoluminescence is a process in which several photons are absorbed, and an electron is excited to a higher energy state with subsequent relaxation and emission of a shorter wavelength photon. Two-photon absorption and luminescence were predicted as early as 1931 \cite{goppert1931uber}, and experimental observations were reported shortly after the development of the first lasers~\cite{kaiser_two-photon_1961}. Three-photon absorption and luminescence processes were observed soon after~\cite{Singh1964}. Today, the applications of two- and three photon excitation processes have matured considerably, and they include three-dimensional microfabrication, data storage, and biological imaging~\cite{Parthenopolos,Kawata2001,Larson,Horton2013}. The efficiency of multiphoton processes in some cases becomes comparable to the efficiency of linear single-photon luminescence \cite{fan2021enhanced}. Compared with single-photon processes, two- and three-photon photoluminescence take advantage of stronger spatial field confinement and weaker Rayleigh scattering. In biological applications, multiphoton processes enable longer penetration depths and less tissue damage. 

Multiphoton luminescence processes of higher orders (with the number of absorbed photons higher than 3)~\cite{chin1995photoluminescence,kim1997multiphoton,ai2021multiphoton} promise further advancements compared to their two- and three-photon counterparts. However, higher-order MPA faces challenges associated with typically much lower multiphoton absorption cross-sections and often lower quantum yield. 

Multiphoton luminescence at the subwavelength scale was studied with various material platforms including plasmonic nanostructures
\cite{Muehlschlegel2005,Wang2007,biagioni_dynamics_2012,ai2021multiphoton}
, rare-earth doped nanocrystals \cite{Qin2004,De2007,Zhao2013,Chen2015,Dong2017,Liao2020,Su2021}, and diamond nanoparticles \cite{Glinka1999}.
Substantial progress has been made in the synthesis of new materials with higher MPA efficiencies, including organic molecules \cite{zheng2013frequency,Correa2007,Hernandez04} and perovskites \cite{chen2017giant,Li19,He2019,He2020,Zhu21}.

Here we conjecture and experimentally demonstrate that the efficiency of higher order multiphoton photoluminescence can be enhanced substantially by employing carefully engineered geometric Mie resonances supported by dielectric subwavelength nanoantennas. High-index resonant dielectric nanostructures provide a new emerging platform for high-field nanophotonics \cite{Carletti2019,Zubyuk,Zograf2022} to complement plasmonic structures \cite{Muehlschlegel2005,Wang2007,biagioni_dynamics_2012,ai2021multiphoton} in a range of applications and functionalities.
The study of geometric resonances in dielectric subwavelength structures has demonstrated a tremendous potential in enhancing nonlinear light-matter interaction, including higher-order nonlinear processes and high-harmonic generation. In this work, we design and fabricate AlGaAs subwavelength particles supporting several Mie resonant modes. We excite the particles in the mid-infrared spectral range and observe five-photon luminescence in the visible spectral range. The interplay of the Mie-resonant modes at the excitation wavelength leads to a suppression of far-field scattering and the enhancement of near-field energy density increasing the overall efficiency of five-photon luminescence. We see a variation of photoluminescence by over 10$^3$ between the resonant and off-resonant cases. Remarkably, the upconversion photoluminescence becomes undetectable outside the spectral range of the resonant mode.

\begin{figure*}[bth!]
 \centering
 \includegraphics[width=0.7\textwidth]{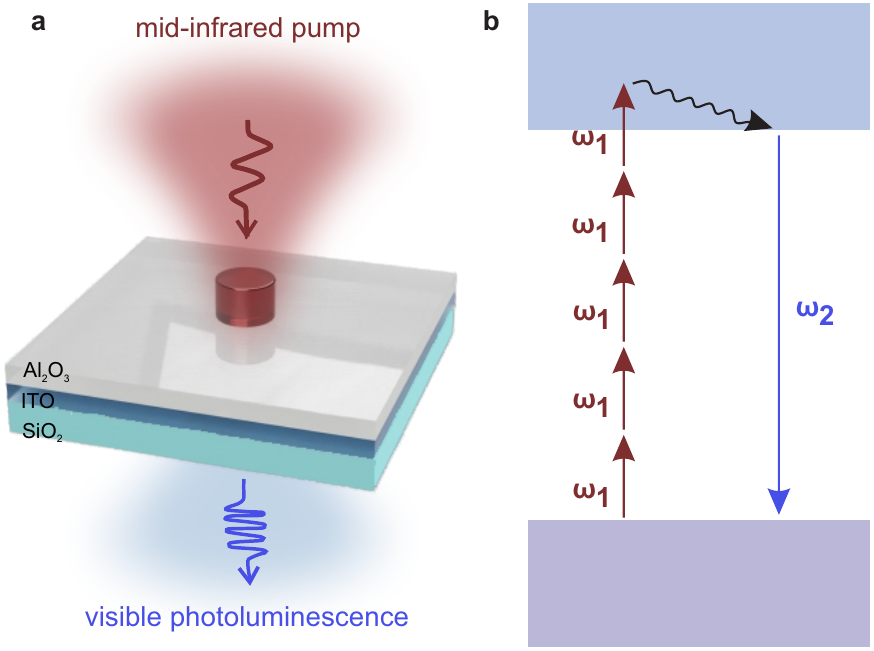}
 \caption{Concept of multiphoton photoluminescence in a subwavelength resonator. (a) Subwavelength AlGaAs resonator emitting visible radiation upon mid-infrared excitation via a five-photon photoluminescence process. (b) Schematic of the five-photon absorption, relaxation, and upconversion emission processes.}
 \label{figure_1}
\end{figure*}

We design and fabricate individual AlGaAs subwavelength resonators optimized for the enhancement of nonlinear light-matter interaction \cite{koshelev2020subwavelength} at the mid-infrared pump wavelength of approximately 3.6 $\mu$m. Figures~\ref{figure_1}a,b present our design and concept of five-photon upconversion luminescence in nanoparticles. 

We fabricate a set of individual resonators from Al(0.2)Ga(0.8)As material with [100] orientation of the crystalline axis using electron beam lithography. The resonators have a fixed height of 1384 nm and diameters in the range of 1300-2200 nm. The material’s bandgap is 1.67 eV, and thus photoluminescence is expected at a wavelength of 740 nm.  The resonators are transferred to a substrate coated with 300 nm indium tin oxide (ITO) and 668 nm-aluminium oxide (Al$_{2}$O$_{3}$) layers with optimized thickness. The layers of Al$_{2}$O$_{3}$ and ITO are employed to increase the performance of the resonator by reducing scattering losses and increase the quality factor ($Q$-factor) of the resonator as discussed earlier~\cite{Carletti2019,koshelev2020subwavelength}.  The resonator sketch and five-photon excitation process are  illustrated schematically in Fig.~\ref{figure_1}.

\begin{figure*}[bth!]
 \centering
 \includegraphics[width=0.99\textwidth]{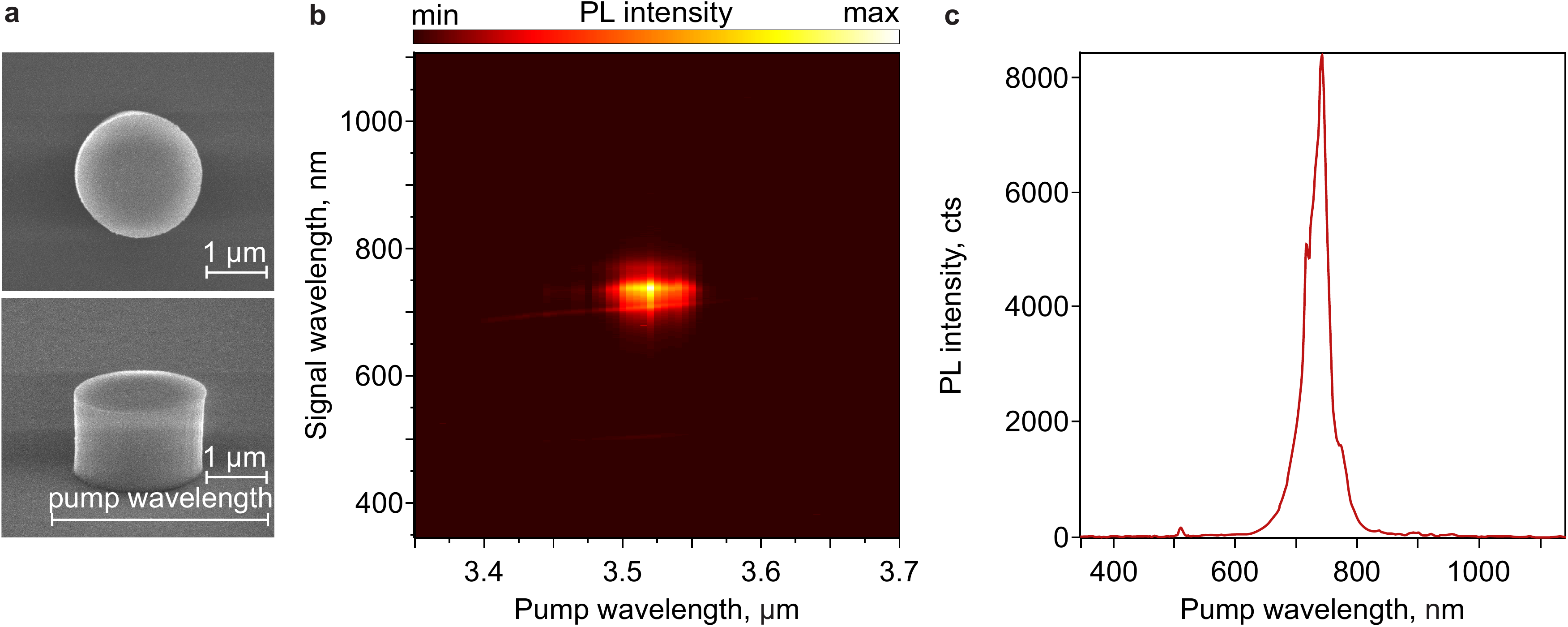}
 \caption{Experimental results. (a) SEM images of the fabricated AlGaAs resonator: top (upper image) and side (lower image) views. (b) Measured signal wavelengths in photoluminescence spectra versus pump wavelengths in a resonator with 1380 nm diameter.  (c) Measured photoluminescence spectrum at a pump wavelength of 3.54~$\mu$m.}
 \label{figure_2}
\end{figure*}

One of the fabricated resonators is shown with the scanning electron microscope (SEM) images in Fig.~\ref{figure_2}a. We pump the resonators in the 3.35 - 3.7~$\mu$m spectral range with a pulsed laser system. The mid-infrared radiation is focused by using an aspheric lens with the numerical aperture of 0.56. The photoluminescence in the visible range is collected in the transmission regime with a Mitutoyo X100 0.7 NA microscope objective and detected with a Peltier-cooled spectrometer (Ocean Optics QEPro). 

Figure~\ref{figure_2}b shows the experimentally recorded spectral scan of the photoluminescence signal of a subwavelength resonator with a diameter of 1380 nm demonstrating a dramatic enhancement at a resonant pump wavelength of 3.53~$\mu$m. The spectra show photoluminescence signals from ~680 nm to 800 nm that peak at ~740 nm. The spectra correspond to the band gap of the Al(0.2)Ga(0.8)As material. Additional weak lines may be seen on the scan in Fig.~\ref{figure_2}b at around 700 nm and 500 nm. They may correspond to the fifth and the seventh optical harmonics generated in the resonator. Remarkably, the photoluminescence signal is detectable only within a narrow spectral range of the excitation at around 3.5 - 3.55 $\mu$m, and the signal vanishes quickly from that excitation wavelength. We attribute this spectral enhancement to a resonant behaviour of the subwavelength particle.  

\begin{figure*}[bth!]
 \centering
 \includegraphics[width=1.0\textwidth]{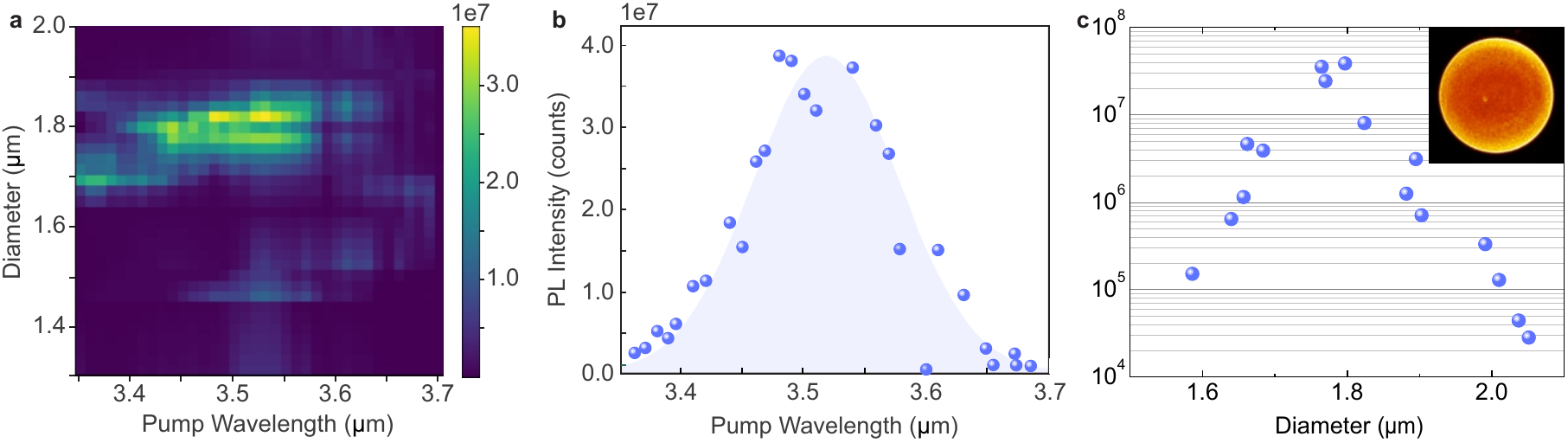}
 \caption{Experimental results. (a) Scan of the photoluminescence signal as a function of resonators' diameters and pump wavelengths. (b) Experimental pump wavelength scan for the optimal diameter of 1800 nm. (c) Experimental diameter scan for  an optimal pump wavelength of 3.48~$\mu$m. Blue shadow area in (b) is a visual guide.  Inset: Directionality diagram of the photoluminescence. The directionality is measured at the back focal plane of the collection objective with numerical aperture of 0.7.}
 \label{figure_3}
\end{figure*}

Next, we study systematically the photoluminescence signal of subwavelength resonators with different diameters (see Fig. \ref{figure_3}). We measure experimentally the dependence of the luminescence intensity on resonators' diameters and pump wavelengths. We extract the maximum values of the spectra for each resonator and pump wavelength, and plot them in Fig.~\ref{figure_3}a.  In the scanned experimental 2D parameter space in Fig.~\ref{figure_3}a, we clearly observe localized enhancements of the photoluminescence signal, which trace the resonant modes. Since the Mie-resonance enhancement is directly related to resonator’s geometric parameters and incident wavelengths, an increase in resonators’ diameters will lead to an increased pump wavelength that produces the maximum photoluminescence intensity. The maximum photoluminescence intensity is observed for the resonator diameter of 1800 nm and wavelength of 3.48~$\mu$m (see Fig.~\ref{figure_3}a). Figure~\ref{figure_3}b shows photoluminescence signal as a function of the excitation wavelength for the resonator diameter of 1800 nm that produces the brightest signal.  Figure~\ref{figure_3}c shows photoluminescence signal as a function of the the resonator diameter for the maximal pump wavelength of 3.48~$\mu$m.  The inset in Fig.~\ref{figure_3}c demonstrates directionality diagram of the photoluminescence signal measured in the back focal plane of our detection system. The photoluminescence signal emission appears approximately isotropic.
Both Fig.~\ref{figure_3}b and Fig.~\ref{figure_3}c feature pronounces local maximum further visualizing resonant enhancement of the photoluminescence efficiency. Figure~\ref{figure_3}c visualizes variation of photoluminescence signal with geometrical parameter within the sensitivity range of our detector.

\begin{figure*}
 \centering
 \includegraphics[width=0.5\textwidth]{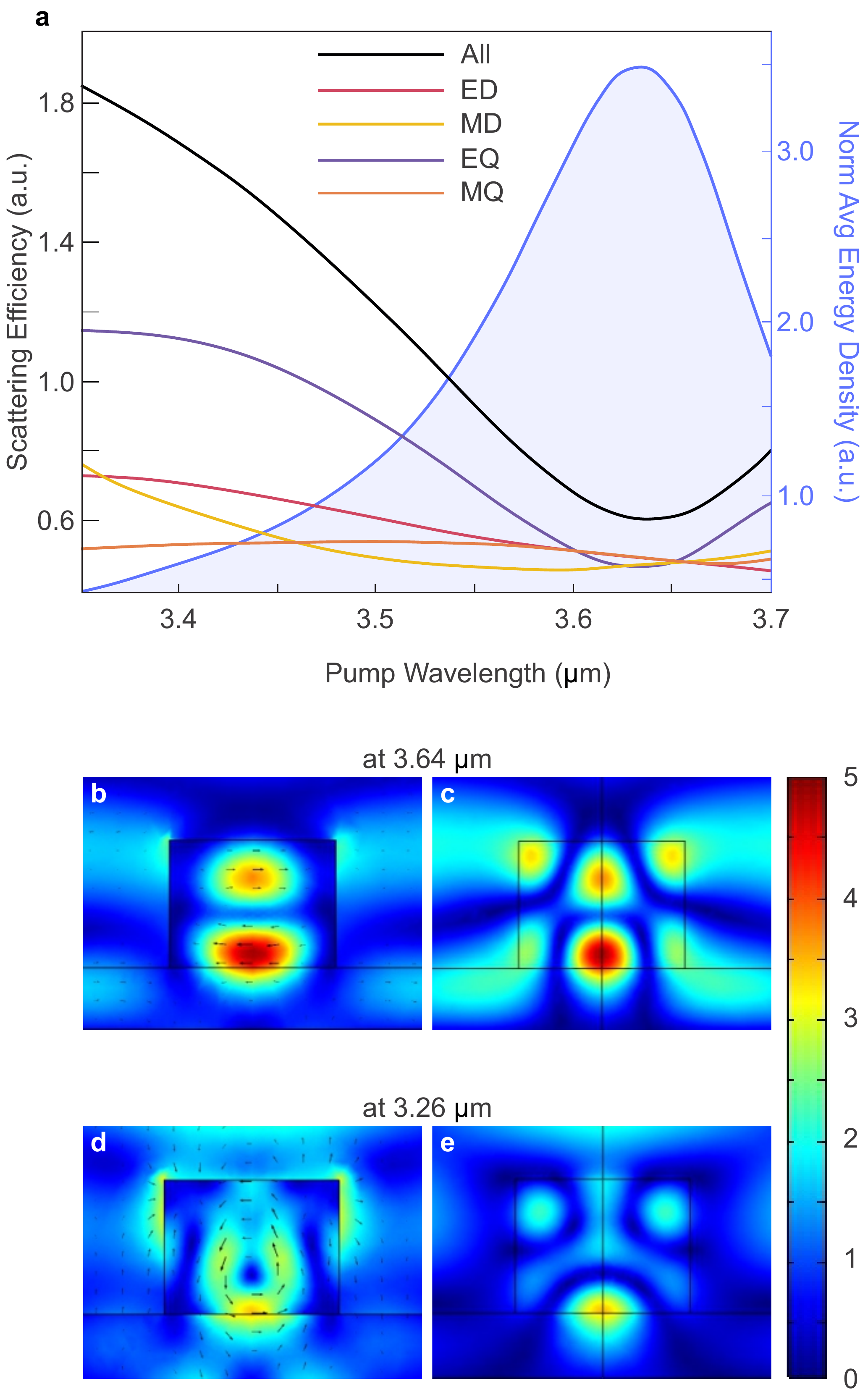}
 \caption{Numerical results for a resonator placed on a substrate with a buried ITO layer. (a) Multipolar response analysis of the scattering efficiency of the resonator with diameter of 1800 nm, which is demonstrated with solid lines of different colors and corresponds to the left axes and the legend. The legend: all - all multipole contributions (black); ED - electric dipole (red), MD - magnetic dipole (yellow), EQ - electric quadrupole (purple), and MQ - magnetic quadrupole (orange). Numerical simulations of the normalized average energy density versus pump wavelength are shown as blue shadowed region and correspond to the right axis. Near-field electric field amplitude distributions at the wavelength of  (b),(c) 3.64~$\mu$m and (d),(e) 3.26~$\mu$m, where (b), (d) $XZ$ plane cut, and (c),(e) $YZ$ plane cut.}
 \label{figure_4}
\end{figure*}

To get a deeper insight into the physics of resonator's response, we perform a decomposition of its total extinction into a series of Mie multipoles (see details in Methods below). We study here the resonator with maximum photoluminescence intensity, which has 1800 nm diameter. Figure~\ref{figure_4}a shows the scattering efficiency spectrum decomposed into the electric and magnetic multipole contributions. The presence of both electric and magnetic Mie resonances is a special feature of Mie-tronics~\cite{mie-tronics}, where the magnetic dipole moment originates from the coupling of the incident light to the circular displacement current of the electric field. This creates strong electric and magnetic field localization and enhances nonlinear processes leading to multiphoton luminescence. Multipolar analysis reveals that the minimum of all multipole scattering contributions (black solid line in Fig. \ref{figure_4}a) corresponds to the pronounced maximum of the averaged energy density (blue shadowed area in Fig.~\ref{figure_4}a), which appears at 3.64~$\mu$m in our calculations. The maximum of the averaged energy density agrees well with the experimentally measured maximum at the photoluminescence spectrum. The difference between the peak of the calculated energy density spectrum and the one of the measured photoluminescence spectrum accounts for about 0.03 $\lambda$- 0.04 $\lambda$. We attribute this spectral enhancement to a resonant behaviour of the subwavelength particle.

\begin{figure*}[bth!]
 \centering
 \includegraphics[width=0.5\textwidth]{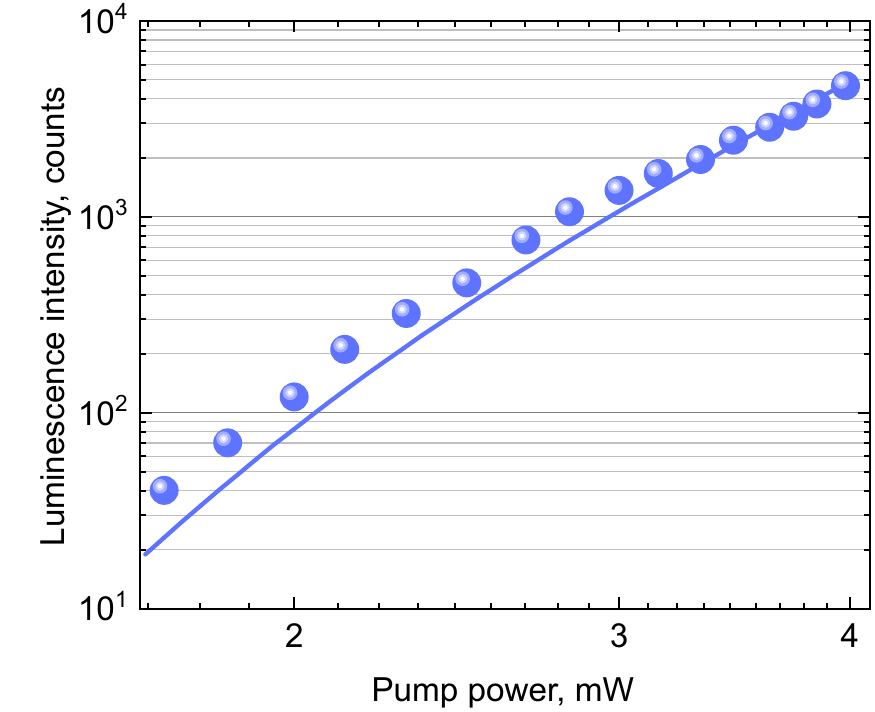}
 \caption{Photoluminescence intensity vs. pump power. Filled circles represent experimental data, solid curve is defined by results of the theoretical model.}
 \label{figure_5}
\end{figure*}

Next, we study experimentally the dependence of the luminescence intensity versus the pump power with the results presented in Fig. \ref{figure_5} (filled circles). We also perform theoretical calculations for the power dependence (Fig. \ref{figure_5}, solid curve).

In order to reach an analytical understanding of the power dependence, we consider a model based on the coexistence of excitons and free carriers in semiconductors~\cite{khmelevskaia2021excitonic}. The ratio between excitons and free carriers before the Mott transition is determined by the Saha-Langmuir equation\cite{szczytko2004determination}:
\begin{equation}
    \frac{x^2}{1-x}=\frac{1}{N} \left( \frac{2 \pi \mu k_b T }{h^2}\right)^{3/2} e^{-(E_b/k_b T)},
\end{equation}
where $x = N_{fc}/N$ is the ratio of free carriers to the total concentration, $\mu = m_e\cdot m_h/(m_e + m_h)$ is the reduced effective mass of the exciton, $k_B$ is Boltzmann's constant, $T$ is the temperature, $E_b$ is the exciton binding energy, and $h$ is Planck's constant. At room temperature, bulk Al$_{0.2}$Ga$_{0.8}$As has a binding energy of approximately 10 meV~\cite{pearah1985low}, so before the Mott transition, free carriers and excitons coexist. The ratio between excitons and free carriers depends on excitation intensity. Further experimental study will be necessary to investigate temporal dynamics of our system. Based on the free carrier and exciton kinetics, the photoluminescence quantum yield can be determined as the ratio of the radiative recombination rate and the total recombination rate:
\begin{equation}
QY=\frac{B N_{fc}^{2}+B^{\prime} N_{ex}}{A N_{fc}+A^{\prime} N_{ex}+B N_{fc}^{2}+B^{\prime} N_{ex}+C N_{fc}^{3}+C^{\prime} N_{ex}^{2}} ,
\end{equation}
 where $N_{fc}$ and $N_{ex}$ are the densities of free carriers and excitons, respectively,  $A$ is a coefficient related to nonradiative
recombination of free carriers on a defect, $B$ is the radiative recombination constant related to spontaneous recombination of free carriers, and $C$ is a constant related to Auger recombination of free carriers; $A'$, $B'$ and $C'$ are coefficients corresponding to the same recombination processes but for excitons. Note that this model neglects the diffusion of charge carriers and the re-circulation of photons.

If the excitation laser pulse is ultrashort, and recombination is negligible during the photoexcitation process ( $\tau_{ex, fc}$ $\sim$ $10^{-10}$ - $10^{-6}$ s),  for five-photon excitation the total charge carrier density can be determined as $N=\eta \gamma I^{5} \tau/E_{ph}$, where $\tau$ is the pulse duration, $\eta$ is the incident light coupling efficiency, $\gamma$ is the nonlinear five-photon absorption coefficient, $E_{ph}$ is the photon energy and $I$ is the intensity of the incident laser pulse. Due to the fact that the carrier concentration  depends on excitation intensity, the quantum yield is
significantly affected by the value of this parameter. Thus, the output photoluminescence intensity can be approximated as:
\begin{equation}
    I_{PL} \sim QY(I)\cdot \gamma I^5.
\end{equation}

Based on these estimates, the power dependence of photoluminescence for a single AlGaAs resonator at room temperature can be calculated by taking into account material parameters \cite{pearah1985low}, and is presented in Fig.~\ref{figure_5} (solid curve). A divergence between the experimental data and calculated curve may be caused by neglecting carriers' diffusion~\cite{zarem1989effect} and photon recycling~\cite{dupont2000efficient} in the model. 
Equation (3) allows us to estimate the field enhancement at the pump wavelength based on the enhancement of the photoluminescence signal. We notice that photoluminescence signal enhancement by the factor of 10$^4$ is expected at an overall local enhancement of pump intensity of approximately just 5 times. This illustrates the impact of subwavelength geometrical resonances on the five-photon photoluminescence signal.

In conclusion, we have observed five-photon luminescence from a single subwavelength AlGaAs resonator. The photoluminescence is visible only in the vicinity of the geometrical resonance of the structure. The experimental results agree with our theoretical calculations based on the five-photon absorption process. Our work paves the way towards the enhancement of efficiency of higher-order multiphoton processes via the physics of Mie resonances.

\section{Methods}

{\bf Numerical Simulations.} The numerical simulations are performed with COMSOL Multiphysics. The dispersion of refractive index and the absorption coefficient are considered for all materials. The AlGaAs values are derived from Ref. \cite{Gehrsitz2000}, for ITO we used ellipsometry data, and for Al$_{2}$O$_{3}$ data is taken from Ref.~\cite{malitson1972refractive}. The physical domains are truncated with perfectly matched layer (PML) domains to avoid spurious reflections. Scattering spectra are calculated from a plane wave excitation with linear polarization from the top of the resonator. The scattering efficiency is defined as
\begin{equation}
    Q_{sca}=\frac{\int_{\Sigma} \mathbf{S}\cdot d \mathbf{n}}{I_0 \pi r^2},
\end{equation}
where the integration is performed on a closed surface, $\Sigma$, including the resonator, $\mathbf{S}$ is the Poynting vector of the scattered field, $\mathbf{n}$ is the normal to the surface $\Sigma$, $I_0$ is the intensity of the incident field, and $r$ is the radius of the AlGaAs nanoresonator.
The average energy density inside the AlGaAs resonators is defined as
\begin{equation}
    \Psi=\frac{\int_V \varepsilon|E|^2 dV}{V}~,
\end{equation}
where $V$ is the AlGaAs resonator volume, $\varepsilon$ is the permittivity, and $E$ is the electric field.
The multipolar coefficients are calculated from the displacement current distribution inside the AlGaAs resonator \cite{Grahn2012}. This is defined as $J(\textbf{r}) = -i\omega\varepsilon_0 \left[\varepsilon_r(\textbf{r})-1\right] E(\textbf{r})$,  where $\textbf{r}$ is the spatial coordinate, $\omega$ is the frequency, $E$ is the electric field, and $\varepsilon_r$ is the relative permittivity. To estimate the electric, $a_E$, and magnetic, $a_M$, coefficients of order $(l,m)$ we follow the method described in Ref. \cite{Grahn2012}. The coefficients for electric dipole (ED), magnetic dipole (MD), electric quadrupole (EQ), and magnetic quadrupole (MQ) contributions are obtained from
\begin{align}
    ED &=  \frac{6}{\pi k_0^2 r^2} |a_E(1,1)|^2~,\\ 
    MD &=  \frac{6}{\pi k_0^2 r^2} |a_M(1,1)|^2~,\\
    EQ &=  \frac{10}{\pi k_0^2 r^2} |a_E(2,1)|^2~,\\ 
    MQ &=  \frac{10}{\pi k_0^2 r^2} |a_M(2,1)|^2~,
\end{align}
where $k_0$ is the wavenumber.

{\bf Fabrication.} The Al(0.2)Ga(0.8)As film with a thickness of 1384 nm is epitaxially grown on a buffer of AlInP layer that is placed on top of a GaAs wafer. A glass substrate with 300 nm-ITO-layer on top of it is commercially purchased. A 700 nm-Al$_{2}$O$_{3}$ spacer is deposited on the ITO/glass substrate using chemical vapor deposition. As an e-beam resist, a polymethyl methacrylate (PMMA) mask is deposited on top of an epitaxially grown AlGaAs film. The top resonator pattern is transferred by exposing it with electron-beam lithography, where the lateral resonator structure is defined using chemically-assisted ion-beam etching. The AlInP layer is removed using wet etching with a diluted HCl solution. Next, the fabricated AlGaAs resonators are attached to a polypropylene carbonate (PPC)-coated polydimethylsiloxane (PDMS) stamp and then transferred to the prepared glass substrate with Al$_{2}$O$_{3}$ and ITO layers on top. The resonators are detached from the PDMS stamp by applying heat (up to 90$^\circ$C) to the thermal adhesive PPC layer. The PPC layer is then removed thoroughly with acetone.

{\bf Optical Experiments.} The measurements are based on a laser system which consists of a
pulsed laser (FemtoLux3 by Ekspla) with a pulse duration of 522 fs, and a repetition rate of 5.14 MHz; and a tunable optical parametric amplifier (MIROPA-fs-M from Hotlight Systems) producing idler radiation in the mid-infrared range from 3.35~$\mu$m to 3.7~$\mu$m with up to 4mW of average power. The pump beam passes through an infrared longpass filter (Edmund Optics).  The mid-infrared radiation is focused with an aspheric lens of 0.56 NA (Thorlabs) onto the sample mounted on a three-dimensional stage. The irradiating output beam is collimated and contracted to match the focusing aspherical lens using a Keplerian telescope based on 50-mm and 200-mm air-spaced achromatic doublets (Thorlabs). The pump beam is observed in reflection by a mid-infrared camera (Tachyon 16k) with a CaF$_{2}$ Plano-Convex lens (Thorlabs). The photoluminescence signal is collected in transmission with a Mitutoyo MPlanApo X100 0.7 NA microscope objective and detected with a Peltier-cooled spectrometer (Ocean Optics QE Pro). Additional plano-convex lenses (Thorlabs) are placed after the collection objective lens for the back focal plane imaging.

\begin{acknowledgement}

L.C. acknowledges support from the Italian Ministry of University and Research (grant 2017MP7F8F). H.-G.P. acknowledges support from the National Research Foundation of Korea (NRF) grant funded by the Korean Government (MSIT) (2021R1A2C3006781) and the Samsung Research Funding and Incubation Center of Samsung Electronics (SRFC-MA2001-01). S.K. acknowledges support from the Australian Research Council (grant DE210100679) and from the European Union's Horizon 2020 research and innovation program under the Marie Sklodowska-Curie grant agreement No.: 896735. Y.K. acknowledges support from the Australian Research Council (grant DP210101292).

\end{acknowledgement}

\bibliography{achemso-demo}

\end{document}